# Generation Model of a Spatially Limited Vortex in a Stratified Unstable Atmosphere

**O.G. Onishchenko**[1,2,*], **S.N. Artekha**[2,**], **F.Z. Feygin**[1,***], **N.M. Astafieva**[2,****]

[1] *Schmidt Institute of Physics of the Earth, Russian Academy of Sciences, Moscow, Russia*

[2] *Space Research Institute, Russian Academy of Sciences, Moscow, Russia*

*e-mail: onish@ifz.ru

**e-mail: sergey.arteha@gmail.com

***e-mail: feygin@ifz.ru

****e-mail: ast@iki.rssi.ru

**Abstract.** This paper presents a new model for the generation of axisymmetric concentrated vortices. The solution of a nonlinear equation for internal gravity waves in an unstable stratified atmosphere is obtained and analyzed within the framework of ideal hydrodynamics. The corresponding expressions describing the dependences on the radius for the radial and vertical velocity components in the inner and outer regions of the vortex include combinations of Bessel functions and modified Bessel functions. The proposed new nonlinear analytical model makes it possible to study the structure and nonlinear dynamics of vortices in the radial and vertical regions. The vortex is limited in height. The maximum vertical velocity component is reached at a certain height. Below this height, radial flows converge towards the axis, and above it, an outflow occurs. The resulting instability in the stratified atmosphere leads to an increase in the radial and vertical velocity components according to the hyperbolic sine law, which turns into exponential growth. The characteristic growth time is determined by the inverse growth rate of the instability. The formation of vortices with finite velocity components, which increase with time, is analyzed. The radial structure of the azimuthal velocity is determined by the structure of the initial perturbation and can change with height. The maximum rotation is reached at a certain height. The growth of the azimuth velocity occurs according to a super-exponential law.

1.    **Introduction**

The existence of vortex structures in the atmosphere is one of the main factors affecting the weather and climate, as a result of the interaction of vortices of different topologies and scales.

In the variety of atmospheric eddy motions, mesoscale and concentrated eddies are clearly distinguished, which are of great interest for both fundamental and applied scientific research. Concentrated eddies (CEs) are nonstationary vertically elongated eddy structures localized in space with a characteristic transverse scale from several meters to several hundreds of meters. CEs includes dust devils (DDs) [Ives, 1947; Balme and Greeley, 2006], more intense and large-scale eddies – tornadoes [Nalivkin, 1983; Justice, 1930], waterspouts [Church et al., 1979], which can be observed on the sea or large lakes, and firespouts, which can suddenly appear during fires in calm weather [Battaglia et al., 2000; Tohidi et al., 2018] or during volcanic eruptions [Thorarinsson and Vonnegut, 1964]. Unlike DDs that carry dust particles, waterspouts involve water droplets in a vertical vortex motion. Despite the fact that the listed vortices arise in different media and can be generated by different natural mechanisms, they all experience an upward helicoidal motion. The rotation velocity in the CEs reaches its maximum value at a certain distance from the vortex axis and tends to zero at its periphery. DDs, as the simplest and most easily observed CEs, are of particular interest for studying the entire class of CEs in the atmospheres of the Earth and Mars.

In analyzing the data of observations of DDs, Sinclair [Sinclair, 1969, 1973] suggested that the necessary conditions for their occurrence are the presence of dust in the near-surface layer of the atmosphere and anomalously high ground temperatures. This is consistent with current models [Balme and Greeley, 2006; Rafkin et al, 2016], in which DDs are formed from convective cells in an unstable near-surface layer with a superadiabatic temperature gradient. A series of observations [Balme and Greeley, 2006; Tohidi et al., 2018] showed that the generation of anticyclonic and cyclonic eddies of this magnitude in open areas is equiprobable. From the observed lack of correlation between external vorticity, generation time, and vortex diameter, it follows that external vorticity alone in the atmosphere is not enough to generate DDs. Meteorological observations [Sinclair, 1973] served as the basis for the development of the first thermodynamic model of DD generation [Renno et al., 2002; Raasch and Franke, 2011]. In this model, warm air in a convectively unstable atmosphere rises and then, as it cools, descends. The proposed model is an analogue of a heat engine that draws energy from a hot surface layer.

Despite a significant amount of previous research, the mechanism of generation and interpretation of the observed vortex structures remains uncertain. In [Onishchenko et al., 2014] the authors proposed a hydrodynamic model of axially symmetric convective vortices (assuming weak perturbations) in a convectively unstable atmosphere at the initial stage of generation. In works such as [Onishchenko et al., 2015; Horton et al., 2016; Onishchenko et al., 2016; Onishchenko et al., 2020], this model was further developed for finite velocity amplitudes in two-dimensional helical motion and various cases of stream functions and seed azimuthal velocities. However, these models were still limited to the analysis of the radial and vertical velocity components of the poloidal motion, either in a very narrow central part or far on the periphery of the convective cell. The aim of this work was to extend the analytical model used to describe the dynamics of an axisymmetric vortex to an arbitrary radial distance from the center. For this, a solution was obtained in the form of Bessel functions (instead of power and exponential functions) using the method of searching for stationary large-scale dipole vortices of Rossby waves in a neutral atmosphere [Larichev and Reznik, 1976].

The structure of the article is as follows. Part 2 derives simplified equations for nonlinear internal gravity waves (IGWs) in an unstable stratified atmosphere. Part 3 discusses the new vortex generation model, and Part 4 explores the proposed model. In the conclusions, the main results of the study are discussed.

**2. Reduced equations**

Meteorological observations served as the basis for creating the first thermodynamic models for the generation of vertical currents (convective cells) [Rafkin et al., 2016; Raasch and Franke, 2011; Renno et al., 1998]. At present, modern ideas about the generation of vertical currents are associated with the instability of the stratified atmosphere. The atmosphere is said to be unstable and stratified if the square of the Brunt – Väisälä frequency or the buoyancy frequency is:

$$\omega_g^2 = g\left(\frac{\gamma_a - 1}{\gamma_a H} + \frac{1}{T}\frac{dT}{dz}\right), \qquad (1)$$

characterizing IGWs as negative. Here, **g** is the acceleration of gravity, $\gamma_a$ is the ratio of the specific heat capacities, $H$ is the local scale of the atmosphere height, and $T$ and $dT/dz$ are the temperature of the medium and the temperature gradient in the vertical direction, respectively. Due to solar heating of the soil, the vertical temperature gradient (the second term of the Brent–Väisälä frequency) is negative, and its value exceeds the first term. The latter corresponds to the well-known Schwarzschild criterion for convective instability. In this case, IGWs turn into unstable exponentially growing cells. When deriving the main equation, we will follow [Onishchenko et al., 2014; Stenflo, 1987]. As the initial system of equations, we used the Euler equation of ideal hydrodynamics (excluding viscosity), which can be written as

$$\frac{d\mathbf{v}}{dt} = -\frac{1}{\rho}\nabla p + \mathbf{g}, \qquad (2)$$

and the transport equation for the potential temperature $\theta$, which is a single-valued function of entropy, which can be written as

$$\frac{d\theta}{dt} = 0, \tag{3}$$

where we neglected dissipative effects (such as thermal conductivity and viscosity). In the equations above, $\rho$ and $p$ denote density and pressure, respectively, $\mathbf{v}$ is the speed of matter, $d/dt = \partial/\partial t + \mathbf{v}\cdot\nabla$ is the Euler (convective) time derivative, $\mathbf{g} = -g\hat{\mathbf{z}}$ is the acceleration of gravity, and $\hat{\mathbf{z}}$ is the unit vector along the vertical axis, $\theta = p^{1/\gamma_a}/\rho$. To close our system of equations, we used the ideal gas equation of state (the Mendeleev–Clapeyron equation) $p/\rho T = const$.

Following the procedure developed in [Renno at al., 1998; Onishchenko at al., 2020; Stenflo 1987, 1990], we can derive a simplified equation for nonlinear IGWs. We introduce a cylindrical coordinate system $(r,\varphi,z)$ with the $z$ axis in the vertical direction and assume that $\partial/\partial\varphi = 0$. In general, the velocity of a divergence-free flow $\mathbf{v} = (v_r, v_\varphi, v_z)$ can be decomposed into its poloidal component $\mathbf{v}_p$ and azimuthal component $v_\varphi \hat{\mathbf{e}}_\varphi$ for $\mathbf{v} = \mathbf{v}_p + v_\varphi \hat{\mathbf{e}}_\varphi$, where $\hat{\mathbf{e}}_\varphi$ is the corresponding unit vector and $\varphi$ is the angle of the cylindrical system. The poloidal velocity components are related to the stream function $\psi(t,r,\varphi,z)$ ratios:

$$v_r = -\frac{1}{r}\frac{\partial \psi}{\partial z}, \quad v_z = \frac{1}{r}\frac{\partial \psi}{\partial r}. \tag{4}$$

According to [Onishchenko at al., 2014] the reduced equation describing the evolution of nonlinear internal gravity waves (IGWs) has the form

$$\left(\frac{\partial^2}{\partial t^2} + \omega_g^2\right)\Delta^*\psi + \frac{1}{r}\frac{\partial}{\partial t}J(\psi,\Delta^*\psi) = 0. \tag{5}$$

Here $J(a,b) = \frac{\partial a}{\partial r}\frac{\partial b}{\partial z} - \frac{\partial a}{\partial z}\frac{\partial b}{\partial r}$ is the Jacobian and the operator $\Delta^*$ defined as

$$\Delta^* = r\frac{\partial}{\partial r}\left(\frac{1}{r}\frac{\partial}{\partial r}\right). \tag{6}$$

The Jacobian in equation (5) corresponds to the so-called vector nonlinearity:

$$J(\psi,\Delta^*\psi) = -\left[\nabla\psi \times \nabla\Delta^*\psi\right]_\varphi.$$

If $\omega_g^2 < 0$, Equation (3) describes the nonlinear dynamics of IGWs in an unstable stratified atmosphere. It is this case that we will consider when, at the moment t = 0 instability occurs, i.e., in (5) we have $\omega_g^2 \to -|\omega_g|^2$. In the opposite case, instability does not arise and the perturbation energy is carried away from the region of their occurrence by IGWs. We note that an equation similar to equation (5) was obtained earlier in [Steflo, 1990] to interpret the behavior of acoustic-gravity vortices.

### 3. Jet generation (radial and vertical flow)

We will be looking for a scalar current function that can generate velocity components, in the form

$$\psi(t,r,z) = v_0 r^2 f(z/L) \text{sh}(\gamma t) \Psi(R), \tag{7}$$

where $v_0 = const$ is some characteristic speed; $\gamma = |\omega_g|$; $R = r/r_0$, $L = const$ is a characteristic spatial scale along the vertical, such that $L \ll H$; $\Psi$ is a function depending on the dimensionless radial distance, $r_0$ is the characteristic vortex radius, and the function $f(z/L)$ will be determined later. Of course, the choice of the stream function in this form is not one-valued, but the function must satisfy the conditions for the regularity of the three components of velocity and pressure on the axis of symmetry of the vortex. For the possibility of an analytical solution, we are looking for a solution by the method of separation of variables. With this flow function, equation (5) reduces to:

$$J(\psi, \Delta^*\psi) = 0. \tag{8}$$

The nonlinear solution of Eq. (8) can be reduced to a linear solution of the form

$$\Delta^*\psi = A\psi, \tag{9}$$

where *A* is a constant. The stream function considered here must remain localized in the radial direction, so it must satisfy the conditions:

$$\left(\psi, \frac{\partial \psi}{\partial r}\right) \to 0, \tag{10}$$

when $r \to 0$ and $r \to \infty$, the function must be regular along the axis of symmetry of the cylinder and vanish at infinity. To find a solution to Eq. (8) that satisfies these boundary conditions, we used the method proposed in [Larichev and Reznik, 1976] for studying large-scale stationary vortices. By applying the operator $\Delta^*$ to the stream function given by equation (7), we get:

$$\Delta^*\psi = v_0 f(z/L)\,\text{sh}(\gamma t)\left(R^2\frac{d^2\Psi}{dR^2} + 3R\frac{d\Psi}{dR}\right). \tag{11}$$

Choosing in equation (9) $A = \pm a_0^2/r_0^2$ and using (11), we obtain the following linear equation for the function $\Psi$:

$$R^2\frac{d^2\Psi}{dR^2} + 3R\frac{d\Psi}{dR} = \pm a_0^2 R^2 \Psi. \tag{12}$$

The general solution of the above equation can be represented as Bessel functions. If $a_0$ is a real number, then with a negative sign on the right, the solution will be a combination of functions: $\Psi(R) = J_1(a_0 R)/R$ and $\Psi(R) = Y_1(a_0 R)/R$. With a positive sign on the right in (12), these can be the following real functions: $\Psi(R) = I_1(a_0 R)/R$ and $\Psi(R) = K_1(a_0 R)/R$. At the zero $a_0$ we get $\Psi(R) = C_0 + C'/R^2$. Since the original equation (5) is nonlinear, the sum of the above solutions is no longer a solution, since the constants in equation (12) on the right will be different. As a result, it is easy to determine that equation (5) will be satisfied identically by each of the above functions, but the ranges of applicability of these solutions in terms of the variable $R$ should differ. Moreover, we see that with this choice the function $f(z/L)$ can be chosen completely arbitrarily.

Among the found solutions, one must choose those that are real and do not give singularities for the velocity components. For the inner region of the vortex, the solutions should lead to zero radial velocity on the axis, and for large distances from the axis the solutions should not oscillate, but rather rapidly decrease. Solutions for different areas, including those for the velocity components, must continuously and smoothly match each other. As well, the solution should be similar to a real concentrated vortex in terms of the ratios of all quantities.

To fulfill conditions (10), we look for a solution to equation (12) by connecting continuous solutions in the internal $\psi_{int}(r < r_1)$ and external $\psi_{ext}(r > r_1)$ areas $r_1$, where the matching of solutions occurs, will be determined later. On the border of the inner and outer regions (with $r = r_1$) the continuity conditions are satisfied:

$$\left(\psi, \frac{\partial\psi}{\partial r}\right)_{int} = \left(\psi, \frac{\partial\psi}{\partial r}\right)_{ext}. \tag{13}$$

In the interior region, all the required conditions are satisfied only by a solution of the following form (we put $a_0 = \delta_0$ at $r < r_1$):

$$\Psi_{int}(R) = \frac{J_1(\delta_0 R)}{R J_1(\delta_0)}. \tag{14}$$

Let us define the characteristic radius $r_0 < r_1$ as the radial distance at which the radial velocity has the maximum absolute value. Then, from the condition $J_0(\delta_0) - J_2(\delta_0) = 0$ we uniquely find the value $\delta_0 \approx 1.841184$. In the outer region, the form of the solution is also unique:

$$\Psi_{ext}(R) = m \frac{K_1(\delta R)}{R K_1(\delta)}. \tag{15}$$

where $m$ and $\delta$ are constants. In order for solutions (7) with functions (14) and (15) to satisfy conditions (13) at the interface $r = r_1$, remaining parameters and value $r_1$ must be related by the following system of equations:

$$\begin{cases} \delta_0 K_1(\delta r_1) J_0(\delta_0 r_1) + \delta K_0(\delta r_1) J_1(\delta_0 r_1) = 0, \\ m K_1(\delta r_1) J_1(\delta_0) = K_1(\delta) J_0(\delta_0 r_1). \end{cases} \tag{16}$$

For illustration, we choose the parameter $\delta = 2$ arbitrarily. Then, from the first equation of system (16) it follows that $r_1 \approx 1.679375$, and from the second equation of system (16) we uniquely obtain $m \approx 2.785071$.

The spatial dependence $\Psi(R)$ and its "smoothness" at the boundary $R = r_1 / r_0$ is shown in Fig. 1 (hereinafter, the values of the parameters found above are used as an example).

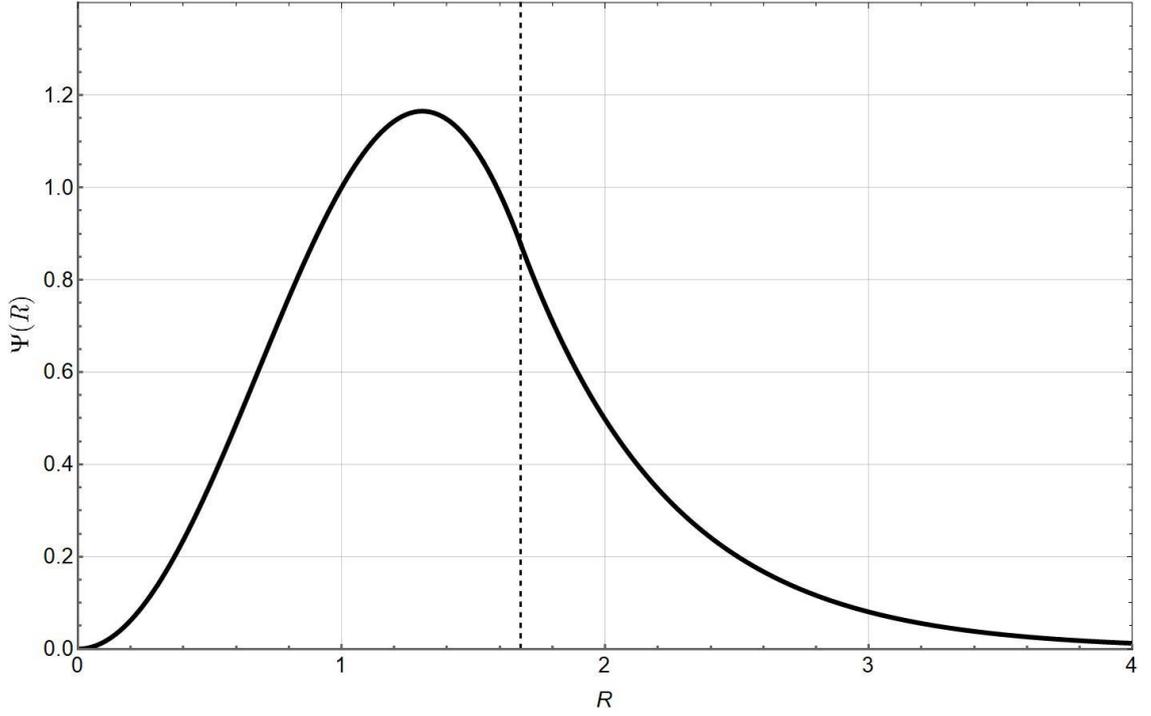

**Fig.1.** The change in the stream function $\Psi(R)$ depending on the dimensionless radial distance $R$.

Then, using equations (4), (7) and (14) or (15), for the radial velocity in the internal $(0 \leq r < r_1)$ and outer areas $(r_1 \leq r < \infty)$ we get the following expressions:

$$v_r^{int} = -v_0 \frac{r_0}{L} f'(Z) \operatorname{sh}(\gamma t) \frac{J_1(\delta_0 R)}{J_1(\delta_0)}, \tag{17}$$

$$v_r^{ext} = -v_0 \frac{r_0}{L} f'(Z) \operatorname{sh}(\gamma t) m \frac{K_1(\delta R)}{K_1(\delta)}. \tag{18}$$

By choosing the type of function $f(Z)$ of the dimensionless parameter $Z = z/L$ we can achieve different z-dependence of the velocity component $v_r$ and $v_z$, for example, for some component (or both) to vanish at the boundaries of the vortex, reach a maximum at a certain height, or change sign starting from a certain height. For now, for simplicity of graphical representation, let us draw the dependence of the radial velocity component for the height where $f'(Z) = 1$, and for the height where $f'(Z) = -1$. For example, with the simplest choice

$$f(z/L) = \begin{cases} (z/L), & 0 \leq z \leq L/2; \\ 1-(z/L), & L/2 < z \leq L, \end{cases} \tag{19}$$

We have $f'(Z)=1$ at $0 \leq z \leq L/2$, an $f'(Z)=-1$ at $L/2 < z \leq L$. The change in the radial component $v_r$ (in units $v_0$) in the inner and outer regions is shown in Fig. 2 for different values of the term $\gamma t$ and specific relationship $r_0/L = 0.1$. This shows that the solutions and their derivatives are continuous at the boundary, and the component $v_r$ is regular on the axis of symmetry and vanishes away from the structure.

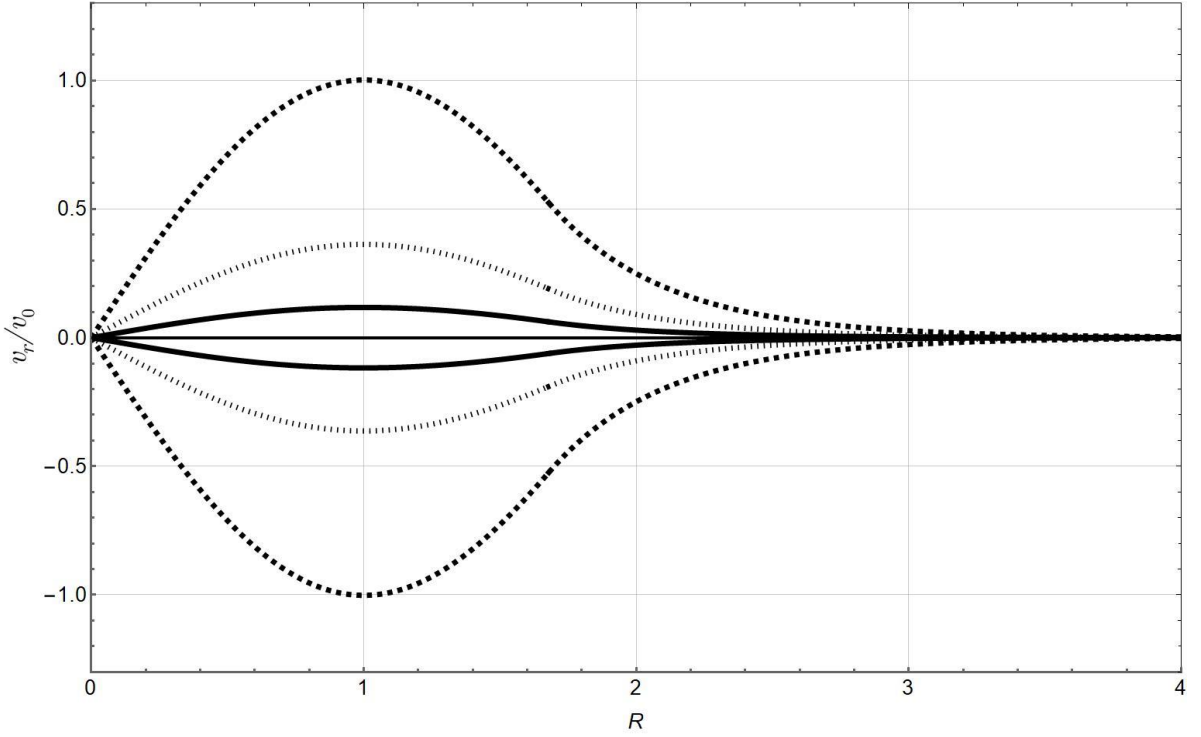

**Fig. 2.** The dependence $v_r(R)/v_0$. The solid, dotted, and dashed lines correspond to the values of $\gamma t$ = 1, 2, 3, respectively; $r_0/L = 0.1$. Negative velocities (inflow) correspond to heights $0 < z < L/2$, positive velocities (outflow) correspond to $L/2 < z < L$.

Similarly, the expressions for the vertical velocity in the inner region ($0 \leq r < r_1$) and outer area ($r_1 \leq r < \infty$) can be written as:

$$v_z^{int} = v_0 f(z/L) \operatorname{sh}(\gamma t) \delta_0 \frac{J_0(\delta_0 R)}{J_1(\delta_0)}, \quad (20)$$

$$v_z^{ext} = -v_0 f(z/L) \operatorname{sh}(\gamma t) m\delta \frac{K_0(\delta R)}{K_1(\delta)}, \quad (21)$$

where the function $f(z/L)$ for the simplest choice is defined in (19). Such a structure of the poloidal fluid motion of convective cells describes vertical flows (or jets) that grow in time. The change in the vertical component of the velocity (in units $v_0$ with respect to the dimensionless radial distance from the center $R = r/r_0$) is shown in Fig. 3 for three increments of $\gamma t$.

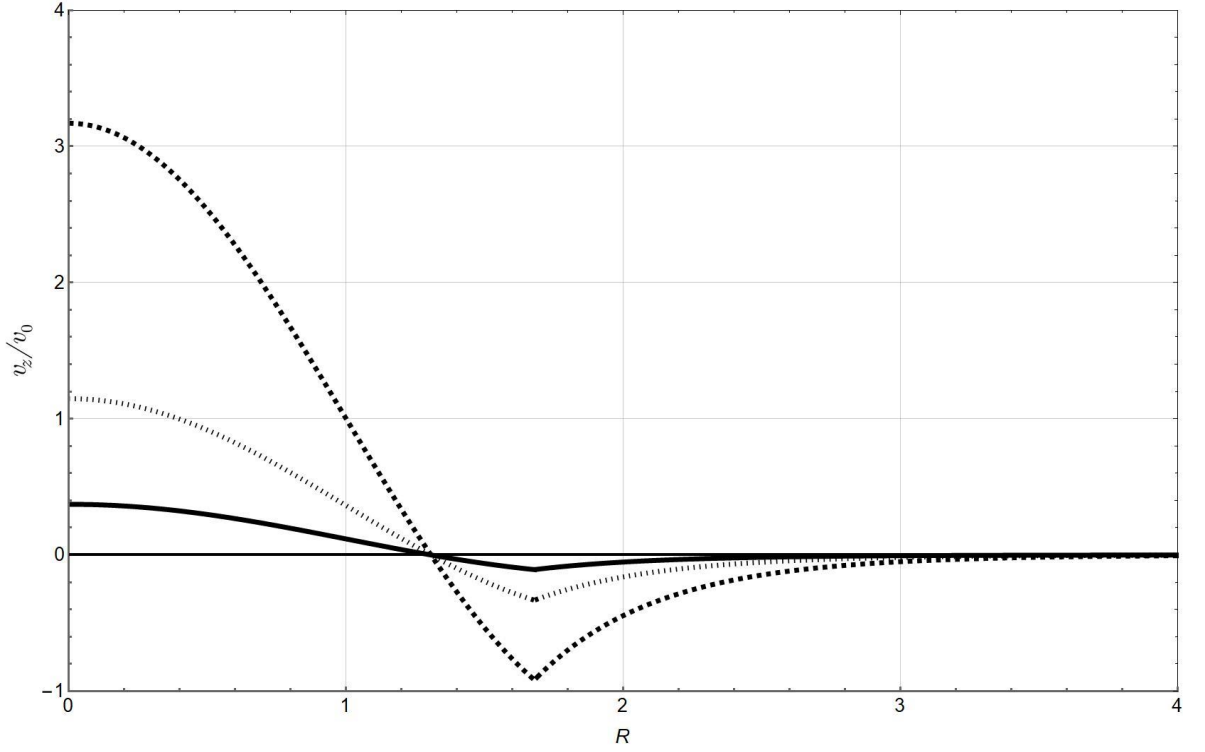

Fig. 3. The dependence of the dimensionless vertical velocity component $v_z/v_0$ from the dimensionless radial distance $R = r/r_0$. The drawing corresponds to the choice $f(z/L) = 0.1$. Solid, dotted and dashed lines correspond to $\gamma t = $ 1, 2, 3. If we choose the simplest dependence (19), then the height $z = L/2$ corresponds to the maximum $z$-velocity component. At this altitude, the corresponding speeds will be 5 times greater than those shown. Further $z$-velocity component decreases with height.

### 4. The vortex generation model

To study the generation of vortex motion, we use the azimuthal component of the momentum equation (taking into account the fact that $\partial/\partial\varphi = 0$):

$$\frac{\partial v_\varphi}{\partial t} + \frac{v_r}{r}\frac{\partial}{\partial r}(rv_\varphi) + v_z \frac{\partial v_\varphi}{\partial z} = 0, \qquad (22)$$

where the radial and vertical velocity components are given by equations (17)–(21). In order to determine the temporal and spatial evolution of the azimuthal velocity component, we will look for a solution for the azimuthal velocity component using the separation of variables method:

$$v_\varphi = v_{\varphi 0} y(t) f_0(z/L) V_{\varphi r}(R), \qquad (23)$$

where $v_{\varphi 0} = const$. In order for the solution of equation (22) to be determined by such a function with separable variables, we obtain the following system of equations:

$$\frac{d y(t)}{dt} = \gamma c_0 \, \text{sh}(\gamma t) \, y(t), \qquad (24)$$

$$\frac{\tilde{v}_r}{Rr_0} + \frac{\tilde{v}_r}{r_0 V_{\varphi r}(R)} \frac{dV_{\varphi r}(R)}{dR} + \frac{\tilde{v}_z}{L f_0(Z)} \frac{d f_0(Z)}{dZ} = -\gamma c_0, \qquad (25)$$

where $c_0 = const$ is some number (a dimensionless constant), and each component with a tilde sign means a multiplier of the same function without time dependence. Then, the solution to equation (24) will be the function:

$$y(t) = \exp\{c_0 (\text{ch}(\gamma t) - 1)\}. \qquad (26)$$

Let us now substitute solutions (17), (18), (20), (21) into equation (25). As a result, we have:

$$f'(Z)\widehat{V}_r(R) + f'(Z) \frac{R \widehat{V}_r(R)}{V_{\varphi r}(R)} \frac{dV_{\varphi r}(R)}{dR} - \frac{f(Z) \widehat{V}_z(R)}{f_0(Z)} \frac{d f_0(Z)}{dZ} = \frac{\gamma c_0 L}{v_0}, \qquad (27)$$

where the notation for the inner $(0 \leq r < r_1)$ and external $(r_1 \leq r < \infty)$ areas:

$$\widehat{V}_r^{int}(R) = \frac{J_1(\delta_0 R)}{R J_1(\delta_0)}, \qquad (28)$$

$$\widehat{V}_r^{ext}(R) = m \frac{K_1(\delta R)}{R K_1(\delta)}, \qquad (29)$$

$$\widehat{V}_z^{int}(R) = \delta_0 \frac{J_0(\delta_0 R)}{J_1(\delta_0)}, \qquad (30)$$

$$\widehat{V}_z^{ext}(R) = m\delta \frac{K_0(\delta R)}{K_1(\delta)}. \qquad (31)$$

It is easy to see that in the simplest case (19) the choice of solution $f_0(Z) = f(Z)$ leads to separation of variables. We note that (19) can be somewhat complicated so that the rate of growth of values along the vertical $C_1$ was arbitrary, and the maximum values were reached not in the middle, but within the vertical dimension of the vortex at an arbitrary point $Z_1$ of the dimensionless variable $Z$: $0 < Z_1 < 1$. We can choose $f(Z) = C_1 Z$, $Z \in [0, Z_1]$, and $f(Z) = C_1 Z_1 (1-Z)/(1-Z_1)$, $Z \in [Z_1, 1]$. In this case, the variables are also separated, the time course of the azimuthal velocity changes, but mathematically this only leads to a change in the constant $c_0$ on the right side of (27), so we consider the simplest case (19). Then, the solution for the remaining radial function, respectively, in the inner region $(0 \leq r < r_1)$ and outer area $(r_1 \leq r < \infty)$ will be:

$$V_{\varphi r}^{int}(R) = \exp\left\{ -\int_R^1 \frac{\alpha_{1,2} + V_z^{int}(x) - V_r^{int}(x)}{R V_r^{int}(x)} dx \right\}, \tag{32}$$

$$V_{\varphi r}^{ext}(R) = \exp\left\{ \int_1^R \frac{\alpha_{1,2} + V_z^{ext}(x) - V_r^{ext}(x)}{R V_r^{ext}(x)} dx \right\}, \tag{33}$$

where the constants $\alpha_1$ and $\alpha_2$ refer respectively to the lower part $0 \leq z \leq L/2$ and to the top $L/2 < z \leq L$ in the height of the vortex, and for these regions the corresponding substitutions are made $c_0 = \frac{\alpha_1 v_0}{\gamma L}$ and $c_0 = -\frac{\alpha_2 v_0}{\gamma L}$. Here, the signs are chosen in such a way that at the initial moment of time the seed azimuth velocity is continuous in height. If the plus sign is chosen for the upper part of the vortex instead of the minus sign this will lead to a very rapid damping of the rotation in the upper half of the vortex. Different values of the constants $\alpha_1$ and $\alpha_2$ correspond to different differential rotations and different dynamics of vortex motion along the height in the region of radial inflow and outflow. Naturally, for the continuity of the flow in the horizontal plane, the quantities $\alpha_1$ and $\alpha_2$ should be the same in the inner area $(0 \leq r < r_1)$ and outer area $(r_1 \leq r < \infty)$ vortex (at the same height). If we choose different values $\alpha_1^{int}$ and $\alpha_1^{ext}$, then this choice corresponds not just to differential rotation, but to rotation with discontinuity (shift at $r = r_1$). Since we are interested in the case of a single vortex with continuous azimuthal rotation, in this case $\alpha_{1,2}^{int} = \alpha_{1,2}^{ext} \equiv \alpha_0$.

Thus, setting the values $\delta, \alpha_0, v_0, v_{\varphi 0}$ and $\gamma$ completely determines the structure and dynamics of the vortex. We will continue the calculations with the previous values of these parameters by selecting $\alpha_0 = 0.01$. We choose for definiteness $/v_{\varphi 0} \mid = v_0$, since the initial speeds (including the initial rotation) must be comparable (and we will look for the ratio of these quantities). From the point of view of physics, this is a question of how much initial vorticity is sufficient for a vortex of this magnitude to arise, and how long the initial swirl must be maintained so that the radial inflow and vertical flow that have arisen due to instability can feed the structure that has arisen. Apparently, from the point of view of ideal hydrodynamics, the swirl must be maintained for a time $\tau \sim 1/\gamma$, until the flow rate increases; then one has to choose $|v_{\varphi 0}| \approx v_0/e$. In addition to the instability resulting in vertical and radial motion, the rotation builds up and is physically maintained by conservation of angular momentum. Therefore, the vortex will grow or be maintained until the entire region with the initial nonzero swirl is pulled up to the axis due to the radial inflow (hence, the value $\alpha_0$ should be very small) and a moving vortex will grow as long as it moves in the region of instability with nonzero swirl (nonzero helicity).

We note that we could choose a different direction of rotation for the upper and lower half of the vortex and a different structure (such a system of coupled cyclonic-anticyclonic rotation is quite common, however, for tropical cyclones). In this case the *R*-distribution of the azimuthal velocity, and its dynamics may differ for the upper and lower half of the vortex.

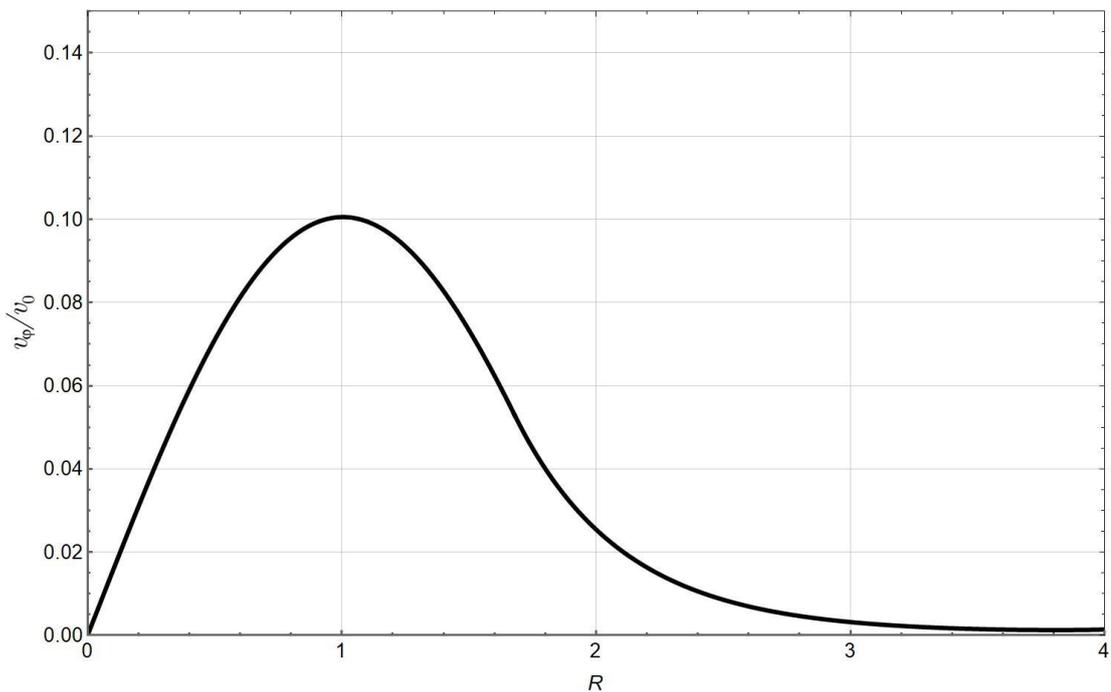

**Fig. 4.** The dependence $v_\varphi/v_0$ for $\gamma t = 1$ on height $z/L = 0.1$.

Figures 2–4 demonstrate the exponential localization of the flow in the radial direction. In particular, Fig. 2 shows the dependence of the radial component of the normalized flow velocity ($v_r/v_0$) on the dimensionless quantity *R* for three different values $\gamma t$ and $r_0/L = 0.1$. The radial

velocity converges on the axis of symmetry and reaches its maximum value at the radial distance $R = 1$. At the initial moment, the velocity is equal to zero, and over time, the growth of the radial component becomes exponential. Figure 3 shows the dependence of the normalized vertical flow velocity $(v_z/v_0)$ from the same dimensionless quantity $R$ at $z/L = 0.1$ and for three different values of $\gamma t$. The vertical speed increases in accordance with equations (20) and (21). It is clear that $v_z/v_0$ reaches its maximum value at the center of the jet. In the area $R \approx 1.3$ the axial velocity component vanishes. In the area $R > 1.3$ the ascending flow in the center of the jet passes into a downward motion and reaches its maximum values at $R = 1.68$. Further, the velocity tends to zero at the vortex boundary.

Figure 4 shows the dependence of the azimuthal velocity component $v_\varphi/v_0$ on the distance $R$. By choosing different values $\alpha_0$ different differential rotations can be obtained. The azimuthal velocity reaches its maximum values at $R = 1$. In the lower half of the vortex at the height maximum ($z = L/2$) the speed will be 5 times greater than that shown on the graph. If the radial and vertical velocities increase by about $e$ times when magnified $\gamma t = 2 \to 3 \to 4 \to \cdots$ per unit (the growth tends to exponential), then the growth of the azimuthal velocity in the lower half of the vortex tends to a super-exponential law. Although at first the azimuthal velocity component increases slowly due to the smallness of $\alpha_0$, but after $\gamma t = 5$ its increase sharply outstrips the growth of the other two velocity components (approximate growth per unit of $\gamma t$: by 1.08, 1.23, 1.77, 4.74, 68.76, 98729, ... times). Therefore, we did not depict this growth.

The solution obtained in this work describes the generation and the initial stage of vortex development; therefore, its applicability is limited in time, since to describe the transition to the quasi-stationary stage and to describe such a stage, it would be necessary to use the Navier–Stokes equations, which significantly complicates the problem.

When $\omega_g^2 > 0$ instability does not arise, all hyperbolic functions in the solutions transform into the corresponding trigonometric functions, perturbations are carried away from the region of origin by IGWs, and the structure does not develop.

### 5. Conclusions

In this work, within the framework of ideal hydrodynamics, a nonlinear equation for IGWs in an unstable stratified atmosphere is obtained, which leads to the formation of axially symmetric structures that grow with time. It is shown that the resulting equation can be reduced to a simpler equation that still contains vector nonlinearity. The stream function $\Psi(R)$ allows us to reduce this nonlinear equation to an equation that has various Bessel functions as solutions. By matching the solutions at the boundary of the convective cell separating the inner region and the outer region, an analytical solution was obtained for the radial and vertical velocity components, which is true for all radial distances $R$.

The proposed model depends on five parameters: $\delta, \alpha_0, v_0, v_{\varphi 0}$ and $\gamma$. This makes it possible to significantly vary the spatial structure of the vortex, including the relative values of the velocity components, the direction of rotation, and the dynamics of the development of the vortex in time. It is obvious that the region of applicability of the solution in time is limited by the initial stage of vortex development (since dissipative processes were not taken into account). For spatial variables, the solution is applicable until the velocities are equal to the background values. Physically, the area of applicability of the solution is limited by the area of existence of instability and the area with nonzero torque.

The solution structure is as follows. The radial velocity converges on the axis of symmetry and reaches its maximum value at a certain distance $R$. From a certain height, the inflow is replaced by a radial outflow. At the initial moment, the radial and vertical velocities are equal to zero; with time their growth becomes exponential. The vertical flow is greatest on the axis of the vortex and reaches its maximum at a certain height. A downward movement occurs along the edges of the vortex.

Thus, the proposed model makes it possible to analyze localized structures of poloidal fluid motion, for example, vertical jets exponentially growing in time depending on $R$. The vortex rotation is differential. The radial structure of the azimuthal velocity is determined by the structure of the initial perturbation. The azimuthal speed can change with altitude. The maximum rotation is reached at a certain height. The growth of the azimuth velocity occurs according to the super exponential law.

The range of applicability of the proposed model is limited to a relatively thin layer of the atmosphere, where convective instability develops, but can be extended to explain the generation of high-speed astrophysical jets or jets in the solar corona.